**Title**:
High-Speed Thermal Imaging of Disk-shaped Firebrands in a Wind Tunnel

**Authors:**
Charles Callahan[a], Erick Gatica[a], Sean Coburn[a], Sam Simons-Wellin[a], Luke Heinzen[b], Laura Shannon[a], John Farnsworth[c], Peter E. Hamlington[a], Gregory B. Rieker[a]

*[a]Department of Mechanical Engineering, University of Colorado, Boulder, 1111 Engineering Drive, Boulder, CO 80309, USA*

*[b]3M Company, 3M Center, St. Paul, MN, 55144-1000, USA*

*[c]Department of Aerospace Engineering, University of Colorado, Boulder, 1111 Engineering Drive, Boulder, CO 80309, USA*

**Abstract**:
The transport and temperature-time history of disk-shaped firebrands was studied with high-speed thermal imaging in a wind tunnel. One millimeter diameter wooden disks were ignited using a novel firebrand generator. The firebrands dropped vertically into a wind tunnel crossflow. Optical images of the firebrands were recorded at 960 frames/second during their transport in the wind tunnel test section at three different wind speeds. The resulting high speed color videos were processed using a particle tracking algorithm and a two color pyrometry technique to determine temperature, position, and velocity time histories of individual firebrands. The firebrand temperatures were observed to oscillate at distinct frequencies that ranged between 50Hz and 480Hz (the Nyquist frequency of the imaging system). A positive correlation was observed between the temperature fluctuation frequency and the relative speed of the firebrand compared to the wind speed. The observed high speed temperature fluctuations suggest that rapid transitions between glowing and flaming combustion may be an important aspect in firebrand transport phenomena and should be considered within the context of model development.

**Introduction:**
Firebrands, also known as embers, are burning fragments of material that are transported by wind. Firebrand transport can facilitate wildland fire spread [1–3], including ignition and destruction of structures in the Wildland-Urban Interface (WUI). Firebrands have been observed to increase the rate of wildland fire spread by up to a factor of three over short ranges [4], and over long ranges by discontinuous spotting [5] . A variety of experimental investigations of the sizes of firebrands produced by the burning of different types of wood under various conditions have been performed by Manzello et al. [6,7], Filkov et al. [2], Adusumilli and Blunck [8] and Bahrani [9]. Many of these experiments show that millimeter scale firebrands make up a significant fraction of firebrands that are generated by different materials such as Ponderosa Pine [8], Douglas Fir [7] and, grass species [9]. This suggests that an accurate understanding of millimeter scale firebrand transport is important for the development of fire spread models.

Several past computational studies present models of firebrand transport [10–16]. However, there are relatively few experimental firebrand transport studies, which are useful for validating computational transport models [17,18]. Tohidi et al. [17] conducted experiments in a wind tunnel that used image processing from recorded videos at 60fps to find a correlation between the maximum rise height and landing location of firebrands. A recent study by Wadhwani et al. [19] validated drag sub-models in a numerical simulation for firebrand transport by accurately predicting the movement of both non-burning and burning firebrands under various conditions. Regarding the temperature analysis of firebrands, Bearinger et al. [20] utilized high-resolution infrared (IR) thermometry based heat flux measurements and inverse heat transfer analysis to study localized heat fluxes from stationary, centimeter scale firebrands. The study revealed that these heat fluxes can be higher than previously estimated. Cantor et al. [21] used controlled experiments with various firebrand types and geometries to develop a standardized fire curve to simulate the thermal effects of firebrand accumulation on dwellings. Manzello et. al. developed a firebrand generator to produce centimeter scale firebrands with a repeatable size and mass distribution [22,23]. These previous experimental studies focus on measuring relatively large firebrands (>1cm) at longer time scales. While larger firebrands are a key driver of medium and long-range fire spotting [18], small firebrands make up a large portion of firebrand production by burning vegetation [7–9] and have been observed to increase the rate of fire spread at short distances [4]. To our knowledge, there have not been experimental studies that focus on small firebrands at short timescales under well controlled conditions.

Here, we seek to create a quantitative benchmark dataset in precisely controlled conditions to inform millimeter scale firebrand transport model development. To be useful for model validation, our experiment aims to have well characterized boundary conditions and track the full state of the firebrands at high spatial and temporal resolution. This is achieved using a novel firebrand generator which introduces disk-shaped firebrands into the top of a wind tunnel. The trajectory, temperature and velocity of the firebrands are measured using a camera which records images at 960 frames per second (fps). The images are processed with a two-color pyrometry technique to extract the firebrand temperature from the images, and a particle tracking algorithm to determine the firebrand position and velocity. The resulting dataset contains the Lagrangian state (temperature, position, velocity) of several thousand disk-shaped firebrands during their lifetime in the wind tunnel test section at three different wind tunnel speeds.

**Experimental Setup:**

**Wind Tunnel:**
The experiments were performed in the WindCline facility at the University of Colorado Boulder [24]. The wind tunnel has test section dimensions 35x35x100cm. Firebrands were dropped from a firebrand generator integrated with the top wall of the wind tunnel approximately 11cm downstream of the test section entrance. Firebrands were observed with a consumer visible wavelength camera until they exited the camera view. The WindCline has the capability to be set to a sloping angle to change the direction of gravity relative to the wind direction, however in the present study the angle was set to zero. More information about the WindCline, including characterization of the flow field, can be found in ref [24].

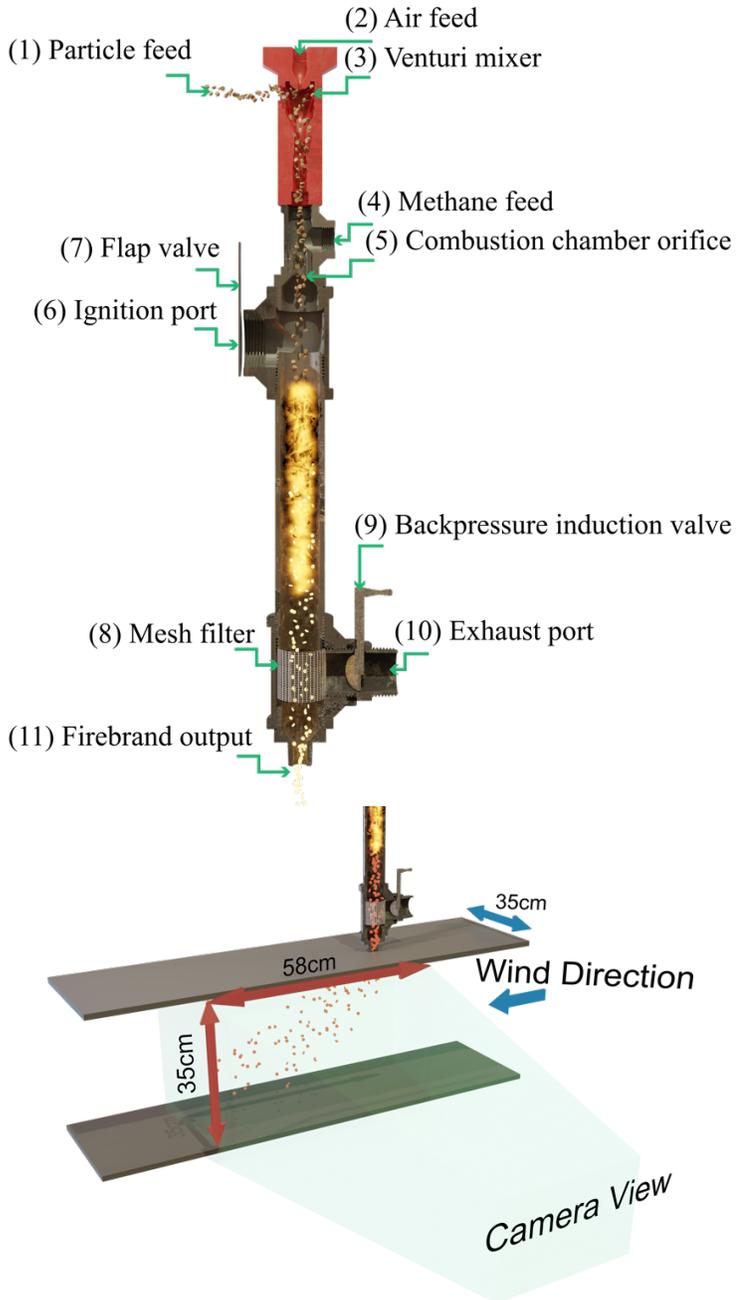

*Figure 1: a) Diagram of the firebrand generator: (1) Particle Feed; (2) Air feed; (3) Venturi mixer; (4) Methane feed; (5) Combustion chamber orifice; (6) Ignition port; (7) Flap valve; (8) Mesh filter; (9) Backpressure induction valve; (10) Exhaust port; (11) Firebrand output. b) An*

*image showing the location of the firebrand dropper and the camera field of view in relation to the wind tunnel test section*

The firebrand generator in this study was designed to ignite firebrands and introduce them into the wind tunnel while minimizing the initial firebrand velocity and hot gas flow from the generator into the wind tunnel cross flow. The firebrand generator is shown in Fig. 1. The wooden disks enter the system at point 1 where they are mixed with the incoming air stream from point 2 in a Venturi mixer (point 3). Methane is added to the air and wood mixture at point 4. The wood, air, and methane mixture then travels into the combustion chamber through a small orifice (point 5). The small orifice increases the velocity of the mixture and reduces the pressure to prevent the mixture from exiting the system through the ignition port (point 6), which is covered with a flap valve (point 7). After ignition through the ignition port during startup, the mixture burns near the middle of the combustion chamber. The methane and air flame rapidly heats the wood disks to their combustion temperature and ignites them. Most of the combustion products then travel through the mesh (point 8), through the backpressure induction valve (point 9) and out of the exhaust port (point 10). The mesh allows the gaseous combustion products to pass through while preventing the firebrands from exiting the system through the exhaust port. The firebrands then fall through a small orifice (point 11) into the wind tunnel. The small orifice at point 11 causes the gaseous combustion products to preferentially exit through the larger exhaust orifice at point 10.

*Backpressure induction valve:*
A filter was installed at the end of the wind tunnel to prevent particulate matter from exiting the tunnel during the firebrand tests. This filter induced a backpressure in the wind tunnel test section of 1-5kPa relative to the atmospheric pressure. To maintain the firebrand flow into the wind tunnel and prevent flow reversal at point 11, the pressure inside of the combustion chamber must not be lower than the pressure inside of the test section. Thus, to increase the pressure in the combustion chamber to match the test section pressure, an adjustable choke valve was added to the exhaust port (point 9) and a flap valve (point 7) was added to the ignition port. The flap valve at point 7 was designed to always be closed except for the moment when the system is ignited. The choke valve can be adjusted to change the pressure difference between the combustion chamber and the test section, changing the flow velocity at the outlet. Thus, the firebrand entry velocity can be adjusted by changing the choke valve position. In these experiments we adjusted the choke valve to minimize the firebrand vertical entry velocity at the highest wind tunnel speed and then maintained the same choke valve position for all experiments.

**Firebrands:**
Wooden disks were formed using a 1mm wooden dowel, the dowel was then sliced into disks using a sharp wood chisel. Ramin hardwood (*Gonystylus spp.)* was used in this study due to its availability in the form of small (1mm) diameter dowels which are used for model ship building. The wood disks can be characterized by their diameter $d$ and their thickness $z$. A jig was used to maintain a consistent $z$ dimension during cutting. After manufacturing a large number of disks for testing, a random sample of ~100 disks was measured to determine the distribution of disk dimensions. The mean disk width, z, was 0.44mm with a standard deviation of 0.06mm and the mean diameter was 1.12mm with a standard deviation of 0.09mm. The distribution of disk diameter and widths are shown in Fig. 1.

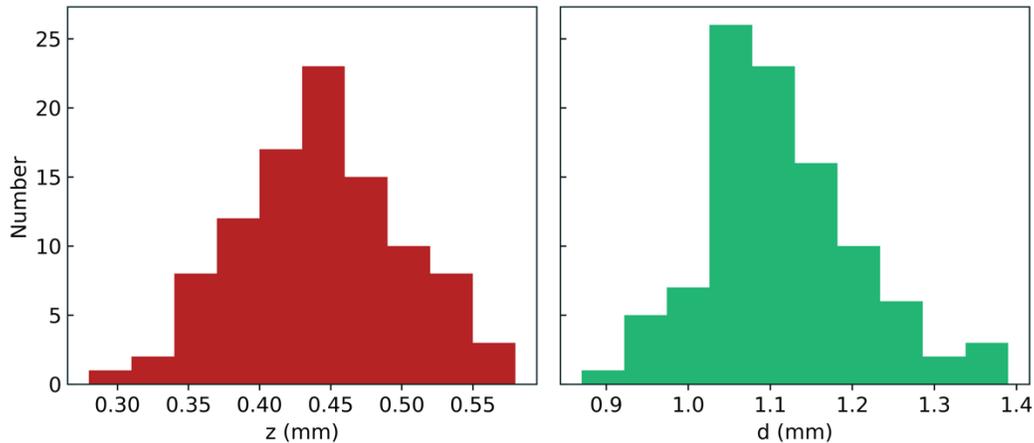

*Figure 2: A sample of the distribution of the width (left) and diameter (right) of the disks used in the experiments*

**Two-color Pyrometry:**

To track the position and temperature of the small firebrands moving quickly through the wind tunnel, a high framerate and high-resolution thermal imaging system is needed. Off the shelf infrared thermal cameras can accurately measure a range of temperatures, however they are expensive and have limited framerates and resolutions due to the use of exotic material detector arrays that often require cooling [25]. Additionally, IR cameras require accurate knowledge of the surface emissivity of the measurement target to generate an accurate temperature. This work employs a two-color visible wavelength pyrometry imaging system based on a consumer grade digital camera (Sony DSC RX10-III). Thermal imaging is achieved by measuring the visible thermal emission from a hot firebrand on two color channels of a digital camera and then using the Planck blackbody emission curve to infer the temperature of the target from the image. This technique has been applied to other high temperature firebrand and soot measurement experiments in past literature [26–34]. Thanks to the maturity of room temperature silicon CMOS image sensors, consumer grade visible wavelength cameras offer very high resolution and framerate at a low cost in comparison to infrared cameras.

Consumer grade digital cameras are composed of a lens that focuses incoming light onto a CMOS image sensor where the spatial distribution of light intensity is digitally recorded. A Bayer filter is included between the lens and the sensor array which contains 3 spectral filters centered around red, green and blue (RGB). The Bayer filter spatially encodes the RGB channels onto groups of adjacent pixels which, after internal decoding by the camera processor, results in a digital array of numbers representing the light intensity for the 3 (RGB) channels at each group of pixels on the array.

If the light entering the camera is emitted by an object at temperature $T$ and with an emissivity as a function of wavelength given by $\epsilon(\lambda)$, the resulting pixel value $C_i$ can be predicted using equation 1.

$$C_i = \int_0^\infty \epsilon(\lambda) I_{bb}(\lambda, T) \eta_i(\lambda) d\lambda. \quad (1)$$

Where $I_{bb}(\lambda, T)$ is the spectral energy density of a black body, given by Planck's law, and $\eta_i(\lambda)$ is the spectral responsivity of the camera of the $ith$ color channel, where $i$ ranges over R,G and B. Making the assumption that the emitter is a grey body, the pixel value ratio $R_{ij}(T)$ of two color channels can be computed independent of emissivity. The grey body assumption has been used routinely in prior literature [35–38].

$$R_{ij}(T) = \frac{\int_0^\infty I_{bb}(\lambda, T) \eta_i(\lambda) d\lambda}{\int_0^\infty I_{bb}(\lambda, T) \eta_j(\lambda) d\lambda} \quad (2)$$

By measuring the camera's spectral responsivity functions for the 3 channels ($\eta_R(\lambda)$, $\eta_G(\lambda)$, $\eta_B(\lambda)$), the ratio for each combination of channels can be simulated using Eq. 2 and one of the ratios can be inverted to compute the temperature.

For this work, the camera's spectral responsivity functions were measured using a monochromator. The spectral responsivity measurement process is detailed in Appendix A. The spectral responsivity and ratios computed from Eq. 2 are shown in Fig. 3 for the camera used in this work.

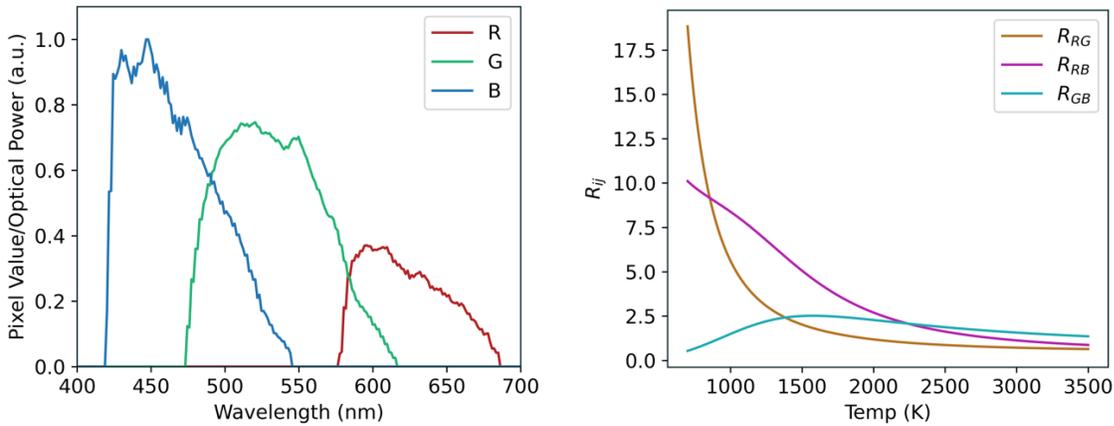

*Figure 3: shows the measured spectral responsivity of the camera (left) and the resulting ratios computed from Eq. 2.*

For this work $R_{RG}$ was used since the R and G channels have a stronger signal compared to the B channel at the expected firebrand temperatures. Additionally, the monotonic decrease of $R_{RG}$ as a function of temperature simplifies the inversion process. The $R_{RG}$ ratio was computed for a range of temperatures up to 3500K and saved for later image processing.

**Data Analysis:**

Eulerian data (mean temperature fields) can be extracted directly from thermal images and videos. Two data processing techniques are required to extract Lagrangian data that represents the state of a single ember during its lifetime in the tunnel test section – a particle finding algorithm to identify individual embers in each video frame, and an interframe particle matching algorithm to temporally match the identified embers between adjacent or closely adjacent video frames.

### Particle finding algorithm

The raw video data is first loaded one frame at a time and for each frame, "particles" are identified and stored for later use during the interframe matching step. Particles are defined as contiguous groups of image pixels which have a greyscale brightness above a threshold.

The process of finding particles in a single video frame starts by computing a mask based on the greyscale threshold. The greyscale threshold is manually chosen to remove background pixels from the processing while keeping pixels representing firebrands. Pixels which are below the greyscale brightness threshold are considered masked and pixels which are above are considered unmasked. After the mask is computed, contiguous islands of unmasked pixels are identified which represent a single firebrand/particle. Average two color pyrometry temperatures are computed for each island of pixels by first averaging the R,G, and B channels for the island, and then using the averaged R and G channels to compute $R_{RG}$ and inverting to get the temperature. Averaging the RGB channels of the particle pixels before computing the temperature has the advantage of increasing the numerical precision of the $R_{RG}$ ratio by $log_2(n_{pixels})$ bits. This is important since the camera outputs only 8 bits of numerical precision per color channel which causes significant bit noise in the resulting $R_{RG}$ ratio and temperature.

The result of the particle finding step is a list of particles which were identified in each video frame, and an average temperature, centroid, and area for each particle.

### Interframe particle matching

After the particles have been identified in each frame, the particles need to be matched between frames to compile the Lagrangian state of each particle during its lifetime in the test section. The relatively simple scheme of assuming that no particle pathways intersect at the same time was used to match particles between frames, due to the low density of embers which were introduced into the test section at a time. The matching algorithm works by first iterating over each particle in each frame and pairing it to the closest particle in the next frame. If no particles are found within a specified radius, the pathway is assumed to stop at that frame. This initial step results in a set of paths where the embers maintain enough brightness to pass through the greyscale filter for consecutive frames. We found that the embers can often become dim for several frames before brightening again, causing the paths to be fragmented. To de-fragment the paths, a post processing step was added to match path fragments together based on the closest spatial and temporal distance between path endpoints. This matching process resulted in all path fragments being assigned to a de-fragmented path (i.e. no particle fragments were removed in this step). Once the particles have been successfully matched between frames, the 2D velocity vector of the particle can be computed using the change in centroid position between frames. The result of this process is a list of particle paths, each containing the particle temperature, centroid position, area, and velocity vector.

**Results and Discussion:**

A series of video segments were captured at each wind condition and concatenated. During each two second video segment, several firebrands were dropped into the test section. The concatenated videos were processed using the particle finding, matching, and thermometry software described in section "Data Analysis". This resulted in a series of particle paths where each path contains a time history of the particle temperature, position, velocity and projected area. The data can be displayed and segmented in a variety of different ways, and for future use the raw particle path data has been included in a json file format. The conditions are summarized in Table 1. The firebrand generator outlet flow speed was determined by introducing sawdust into the firebrand generator, then measuring the velocity of the resulting particles at the firebrand generator outlet using the imaging system. Sawdust was used since it produces much smaller firebrand particles whose motion is primarily driven by the velocity of the surrounding gas.

| Condition Number | Wind speed (m/s) | Firebrand generator outlet flow speed (m/s) |
|---|---|---|
| 1 | 2.0 | 1.9 |
| 2 | 1.1 | 1.3 |
| 3 | 0.4 | 1.6 |

*Table 1 shows the wind speed and inlet flow speed at each condition.*

**Spatial distribution**

Figure 4 shows the temperature time-history and path traveled by the particles measured during the three conditions. As expected, the average particle paths tend to travel over a larger x distance at higher wind speeds. Additionally, temperature "flickering" can be observed during particle travel. These images were produced by redrawing the integrated particle path data onto a single image using the processed temperature and position information.

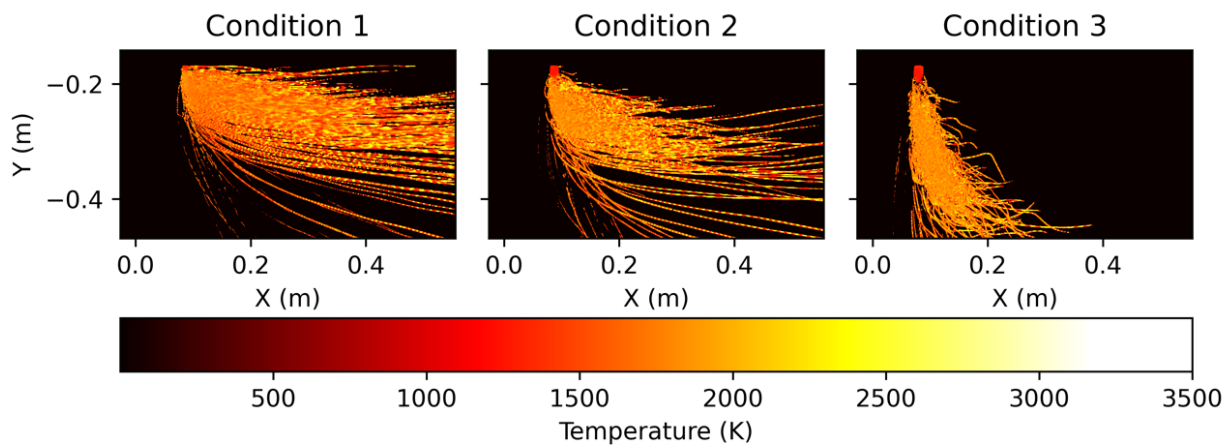

*Figure 4: shows the aggregate temperature time history and path traveled by the particles measured during the three conditions.*

### Temperature time history and distribution

The flickering effect that was observed in Figure 4 can be better visualized using the Lagrangian state of a single particle. An example of the temperature time history of a single particle is shown in Fig. 5, where a distinct oscillation is apparent. Fig. 5 also shows that the oscillation frequency seems to decrease as time progresses.

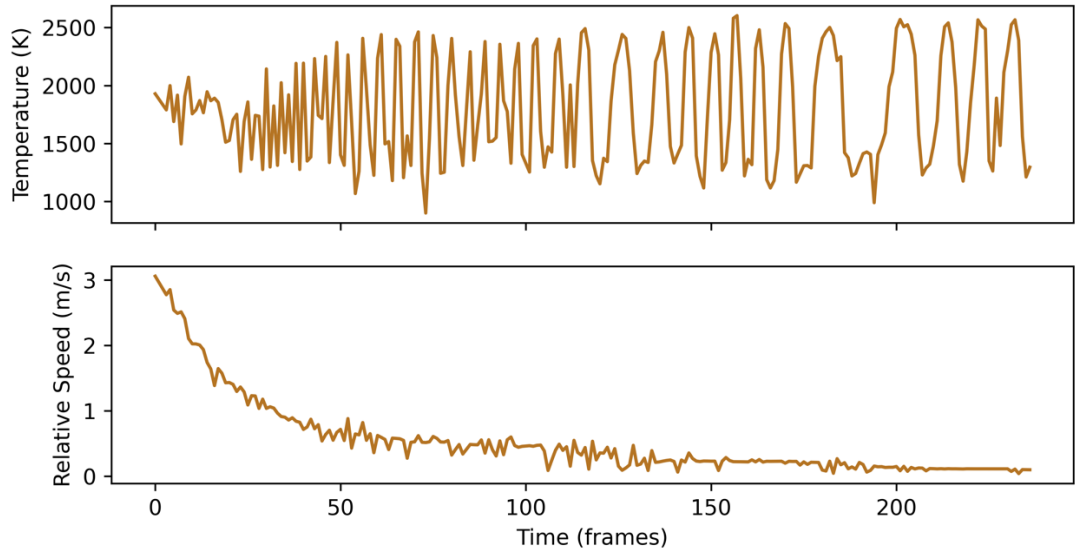

*Figure 5 shows an example of the temperature oscillation of a single firebrand from Condition 1 during its lifetime in the test section (top) and the speed of the firebrand relative to the wind (bottom).*

This oscillatory behavior is apparent in the temperature time history of many of the particles. To quantify the prevalence of the oscillation over the ensemble of firebrands that were measured, temperature distributions can be computed from the particle path data by assembling a set of temperatures from the state of each particle during each frame. The resulting sets represent the temperature distribution of all particles at all times at a particular wind condition. The temperature distributions are shown in Fig. 6. The results show a bimodal temperature distribution which becomes less bimodal at lower wind speeds. The bimodal distribution can be attributed to the temperature oscillation that is evident when viewing the Lagrangian state of each firebrand particle. The two peaks of the bimodal distribution correspond to the minimum and maximum temperature during the oscillation. The strong bimodal distributions show that the temperature oscillations are present in a large portion of the measured firebrands.

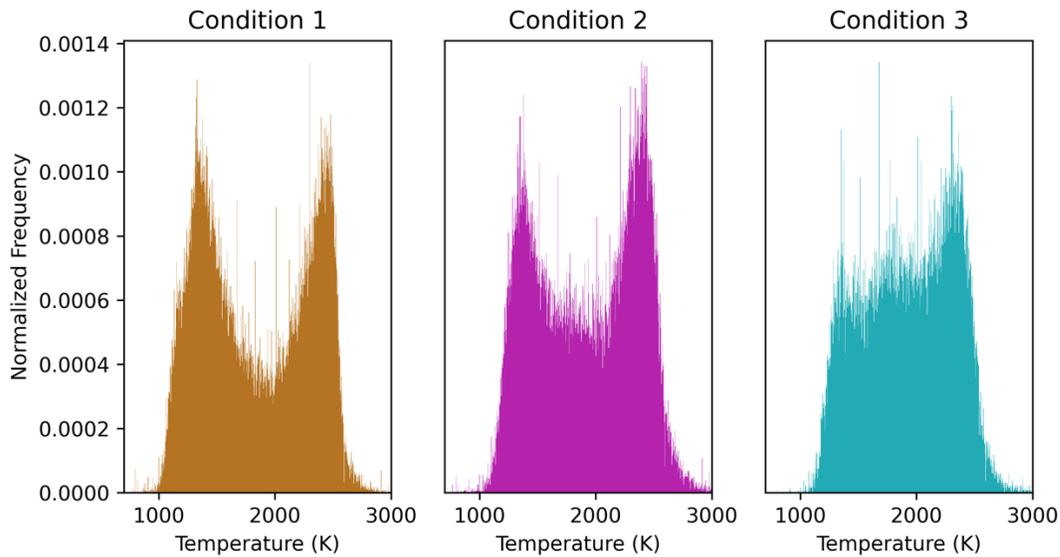

*Figure 6: shows the distribution of measured firebrand temperatures during the 3 conditions.*

**Temperature fluctuation correlations**

This section analyzes the temperature oscillations evidenced by the bimodal temperature distribution shown in Fig. 6 and the single particle temperature histories such as Fig. 5. Particle temperature oscillation frequencies were computed by fragmenting the particle time histories into segments representing 50 video frames (or 0.052 seconds) and computing the Fast Fourier Transform (FFT) of each segment. The frequency with the maximum amplitude in each segment was recorded (the 'peak frequency'). Note that the first two points of the FFT were removed from the analysis to remove the large signal representing the constant temperature offset. A "weight" factor was then computed by dividing the amplitude of the spectrum at the peak frequency by the mean amplitude of the rest of the spectrum. This weight factor is large when the segment has a well-defined peak fluctuation frequency and small if not. The weight factor is used in later analyses to compute weighted averages. In addition to the fluctuation frequency, the mean velocity vector over each 50-frame time segment was recorded. The result of this analysis was a time series of particle temperature oscillation frequency and mean velocity vs time.

The correlations between frequency and relative speed were then computed by segmenting the relative speed and frequency data for all of the measured particles into 30 equally spaced relative speed bins and then computing the weighted average frequency in each bin using the pre-computed weight factor. The results are shown in Fig. 6. The results show that the temperature oscillation frequency increases with the speed of the particle relative to the wind.

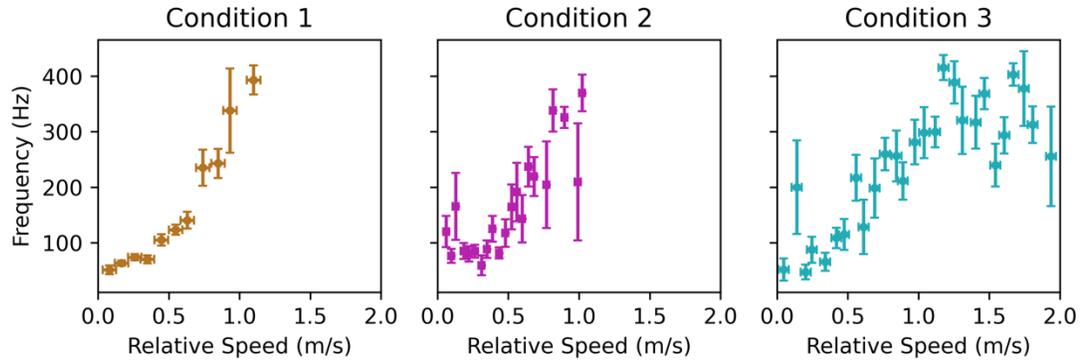

*Figure 6: Shows the correlation between temperature oscillation frequency and the speed of the firebrand particle relative to the wind which was measured during the three runs. The correlations represent the ensemble of all of the measured particles. The error bars on the x-axis represent the speed range of the averaging bin and the y-axis error bars represent the weighted standard error of the set of fluctuation frequencies in each bin.*

The correlation between relative wind speed and temperature oscillation frequency suggests that the oscillations are driven by interactions between the fluid and the particle. One possibility is that the relative wind speed changes the rate at which combustion products and ash (which could slow the combustion reaction) are removed from the particle surface. Another possibility is that oscillatory fluid behavior such as vortex shedding rapidly drives the transition between glowing and flaming combustion. Additionally, the oscillatory behavior could be a result of the firebrand disks tumbling and/or rotating in the air flow, changing cross section and the dynamics of the supply of reactants to the combustion reaction. Additional computational and higher resolution experimental data are being pursued to determine the exact cause of the oscillations.

**Conclusion:**

We present an experimental study on millimeter scale firebrand transport in a wind tunnel. Millimeter diameter wooden disks were manufactured and a novel firebrand generator was used to ignite the disks and drop them into a wind tunnel. Images of the disks were taken at 960 fps using a consumer grade color digital camera, and a two color pyrometry technique was used to convert the image pixel values to temperatures. The image data were further processed using a particle tracking algorithm to identify and match particles between frames so that Lagrangian firebrand states could be recorded. The results showed that the firebrand emission temperature oscillated at up to the Nyquist frequency of our imaging system (480Hz), producing a bimodal temperature distribution through time. A further analysis of the oscillation frequency was performed which showed that the oscillation frequency increased with the relative speed of the firebrands with respect to the wind. The results of this experimental study suggest that fast transient dynamics may play an important role in accurately modeling the behavior of millimeter scale firebrands.

**Acknowledgements:**

This work was supported by the Strategic Environmental Research and Development Program (Grant No. W912HQ-20-C-0065), and 3M Corporation.

**Camera calibration (Appendix A)**

To develop the simulated pixel ratios $R_{RG}, R_{RB}, R_{GB}$, the spectral responsivity of the three color channels ($\eta_R(\lambda), \eta_G(\lambda), \eta_B(\lambda)$) of the camera must first be measured. The spectral responsivity function describes the pixel value that is produced per unit spectral power incident on the sensor array.

    A monochromator was used to generate a narrow bandwidth wavelength tunable light source. The light source was composed of a tungsten lamp and a grating for spectral separation. The grating output was blocked except for a small slit to allow a narrow bandwidth of light to pass through the output. The grating was then rotated to set the output wavelength.

    First the monochromator output spectral power (P) vs grating angle θ was measured by rotating the grating and measuring the power output using a Thorlabs PM100D power meter with an S140C power head. Then, to measure the camera pixel response functions, the monochromator output was sent into the camera lens and focused onto ~50 pixels near the center of the camera image sensor. The monochromator was then scanned through its wavelength range and the pixel values which were illuminated by the monochromator output were recorded. The illuminated pixels were averaged together and indexed to the grating angle to result in the pixel values vs grating angle $V_R(\theta), V_G(\theta) \ and \ V_B(\theta)$. Finally the resulting pixel response functions were computed by dividing the pixel values by the monochromator output power and remapping the angle (θ) to the output wavelength λ.

**Furnace Validation**

    To validate the temperature inversion process and estimate the uncertainty of the two color pyrometry system, we measured the temperature of the tip of a hot thermocouple using the camera. The thermocouple was placed inside of a tube furnace and the camera was set up to image the tip of the thermocouple. Then the resulting two color pyrometry measurement was compared to the thermocouple temperature over a the range of temperatures up to the maximum temperature of the furnace. The resulting validation measurement is shown in Fig. 7. The results show good agreement between the two temperature sensors, with an average error of 53K.

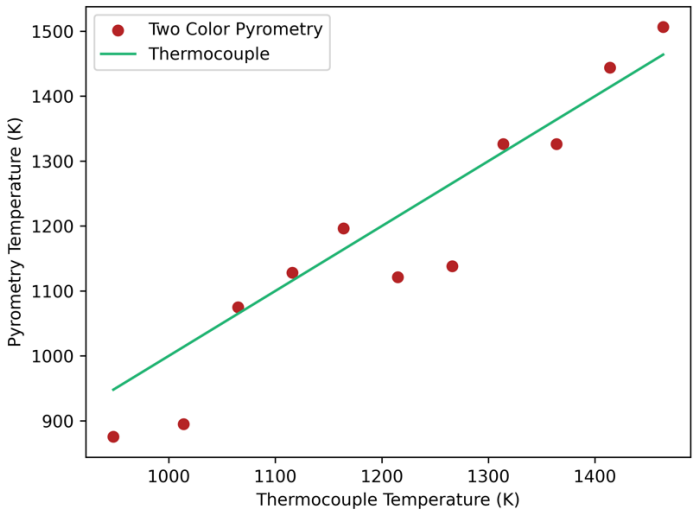

*Figure 7: shows the furnace validation measurement. The X axis represents the furnace temperature measured with a thermocouple, and the y axis shows the temperature which was measured by imaging the tip of the thermocouple with the pyrometry imaging system*

**Open Source data format (Appendix B)**

The particle data is provided so that it can be used for computational validation and model development. We include the post processed particle path data due to the large file size of the original videos. The data is stored in three json (JavaScript Object Notation) text files. The files contain a list of interframe particle IDs which each represent a single Lagrangian particle time history. Each interframe particle ID contains a list of frame numbers which each contain a table of the particle state during that frame. Each particle state table contains the following fields (shown in Table YY). Note that the framerate is 960 frames/s for each dataset.

| Field | Description | Units |
| --- | --- | --- |
| centroidX | Centroid position (X) | Pixels |
| centroidY | Centroid position (Y) | Pixels |
| velocityX | Velocity (X) | Pixels |
| velocityY | Velocity (Y) | Pixels |
| nPixels | Projected Area | Pixels$^2$ |
| temp | Temperature | Kelvin |

*Table YY shows the state information for a single particle measurement.*

Additionally, we include a file called rundata.json for each run which contains the information required to convert pixels to physical distance and the wind velocity for each run. The rundata fields are shown in Table YY.

| Field | Description | Units |
| --- | --- | --- |

| originX | Position of the firebrand outlet (X) | Pixels |
|---------|--------------------------------------|--------|
| originY | Position of the firebrand outlet (Y) | Pixels |
| scaleX | Image scale (X) | Pixels/cm |
| scaleY | Image scale (Y) | Pixels/cm |
| windVx | Wind tunnel velocity X | m/s |
| windVy | Wind tunnel velocity Y | m/s |

*Table YY shows the rundata.json file format.*